\documentclass{svjour2}                    
\smartqed  
\usepackage{graphicx}
\def\sax{{\it Beppo}SAX}

\def\amx{{$\alpha_{\mu x}$~}}
\def\amg{{$\alpha_{\mu \gamma}$~}}
\def\amgB{{$\alpha_{{\mu \gamma}_{100\%CGB}}=0.994$~}}

\def\ax{{$\alpha_{\rm x}$~}}
\def\ergs{{erg~cm$^{-2}$s$^{-1}$~}}
\def\phs{{ph~cm$^{-2}$s$^{-1}$~}}
\def\ergssd{{erg~cm$^{-2}$s$^{-1}$deg$^{-1}$~}}
\def\ergj{{erg~cm$^{-2}$s$^{-1}$Jy$^{-1}$~}}
\newcommand{\lsim}{{\lower.5ex\hbox{$\; \buildrel < \over \sim \;$}}}
\newcommand{\gsim}{{\lower.5ex\hbox{$\; \buildrel > \over \sim \;$}}}

\def\fxfr{$f_{\rm x}/f_{\rm r}$~}
\def\nupeak{$\nu_{peak}$~}

\def\ee{\end{equation}}
\def\be{\begin{equation}}

\begin{document}

\title{Non-thermal Cosmic Backgrounds and prospects for future high-energy observations of blazars
}


\author{P. Giommi         \and
        S. Colafrancesco 
}


\institute{P. Giommi \at
              ASI-ASDC \\
              Tel.: +39-06-94188870\\
              Fax: +39-06-94188870\\
              \email{paolo.giommi@asi.it}           
           \and
           S. Colafrancesco \at
              INAF - Osservatorio Astronomico di Roma \\
              Tel.: +39-06-94286418\\
              Fax: +39-06-9447243\\
              \email{Sergio.Colafrancesco@mporzio.astro.it}           
}

\date{Received: date / Accepted: date}

\maketitle

\begin{abstract}
We discuss the contribution of the blazar population to the extragalactic background
radiation across the electromagnetic (e.m.) spectrum with particular reference to the
microwave, hard-X-ray and $\gamma$-ray bands. Our estimates are based on a recently
derived blazar radio LogN-LogS  that was built by combining several radio and
multi-frequency surveys.  We show that blazar emission integrated over cosmic time gives
rise to a considerable broad-band non-thermal cosmic background that dominates
the extragalactic brightness in the high-energy part of the e.m. spectrum. We also estimate the
number of blazars that are expected to be detected by future planned or hypothetical
missions operating in the X-ray and $\gamma$-ray energy bands.
\end{abstract}

\section{Introduction}
\label{intro}

Active Galactic Nuclei (AGN) are well known to dominate the high-energy (soft-X-ray and beyond) high Galactic latitude sky. Their radiation integrated over cosmic time can explain most, if not all, the extragalactic background radiation (e.g. \cite{Giacconi62,Rosati02,Moretti03,Giommi06}). Historically, AGN have been classified in many, sometime inconsistent or confusing. ways, depending on how they appeared in surveys performed in different energy bands and flux limits, often determined by the technology available at the time of the survey, or depending on some observational parameter like e.g. the flux ratio between radio and optical emission, the equivalent width of their emission lines, etc.

In the following we use the widely accepted standard paradigm where AGN are powered by accretion onto a super-massive black hole, e.g. \cite{Urry95,Urry2003,Vellieux2003}, to divide this class of sources. into two broad categories defined by their intrinsic emission mechanism.

\begin{itemize}
\item {\bf AGN whose power is dominated by radiation of thermal origin} produced by matter that is strongly heated in the inner parts of a disk of matter falling onto the super-massive black hole and illuminates the circumnuclear matter that is responsible for the broad line emission.
We call these objets Thermal Emission Dominated AGN  or TED-AGN.

\item  {\bf AGN whose power is dominated by Non-Thermal radiation} (or NTED-AGN) where most of the emission is generated through non-thermal processes like the synchrotron and the inverse Compton mechanism by particles accelerated in a jet of material that moves at relativistic speed away from the central black hole. The jet itself if formed converting part of the accretion energy  in a way that is presently not well understood.

\end{itemize}
Within this general definition the class of blazars corresponds to the small subset of
NTED-AGN that are viewed at a small angle w.r.t. the jet axis (for this reason their
emission is strongly amplified by relativistic effects \cite{bla78,Urry95}), whereas
Radio Galaxies are those NTED-AGN that are viewed at a large angle w.r.t. the jet axis.
Here we do not distinguish between line-less objects, or BL Lacs, and broad-line
Flat-Spectrum Radio Quasars, or FSRQs.

When the accretion and jet emission coexist in the same object and produce similar
amounts of radiation we have AGN that are a mixture of the TED and NTED type that
appear like broad-line radio galaxies (e.g. 3c120).

The overall cosmic background energy has two well understood components: the primordial
black body emission peaking at microwave frequencies, or CMB, and the X-ray apparently
diffuse emission arising from the accretion onto super-massive black holes in AGN
integrated over cosmic time, or CXB. We will show that blazars add a third non-thermal
component that at low frequencies contaminates the CMB fluctuation spectrum and
complicates its interpretation \cite{giocol04}, while at the opposite end of the e.m. spectrum
it dominates the extragalactic background radiation \cite{Giommi06}.

In the following we estimate the contribution of blazars to non-thermal cosmic backgrounds starting from a recently derived deep radio LogN-LogS. The broad-band electromagnetic spectrum of a blazar is composed of a synchrotron low-energy component that peaks [in a $Log(\nu)-Log(\nu f(\nu))$
representation] between the far infrared and the X-ray band, followed by an Inverse
Compton component that has its maximum in the hard X-ray band or at higher energies,
depending on the location of the synchrotron peak, and extends into the $\gamma$-ray or
even the TeV band. Those blazars where the synchrotron peak is located at low energy are
usually called Low energy peaked blazars or LBL, while those where the synchrotron
component reaches the X-ray band are called High energy peaked blazars or HBL \cite{P95}. LBL sources are the large majority among blazars
\cite{Pad03} and are usually discovered in radio surveys, while HBL objects are
preferentially found in X-ray flux limited surveys, since at these frequencies they are
hundreds, or even thousands, of times brighter than LBLs of similar radio power.

\section{The radio LogN-LogS of blazars and their contribution to extragalactic cosmic backgrounds}
\label{sec:lognlogs}

The deep blazar radio (5GHz)  LogN-LogS shown in Fig. \ref{logns} has been recently
derived  by \cite{Giommi06} who showed that it can be described by a broken power law
with parameters defined in eq. (\ref{eq.L1}):

\begin{figure}[ht]
\centerline{
\includegraphics[width=5.cm, angle=-90] {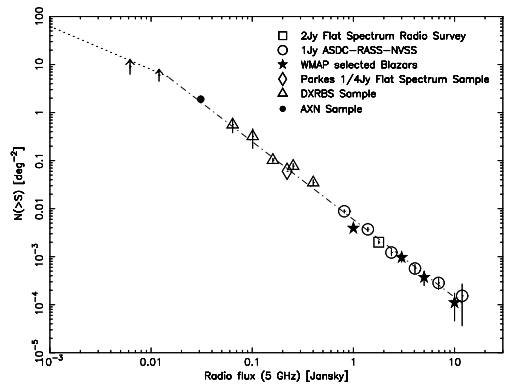}
} \caption{The blazars radio (5GHz) LogN-LogS of  \cite{Giommi06} who built it combining several  surveys as indicated in the top right part of the plot (see \cite{Giommi06} for details).
}
 \label{logns}
\end{figure}

\begin{equation}
\displaystyle {N(>S)} =
 \cases{
 5.95~10^{-3}\times S^{-1.62} &  $S > 0.015~Jy$ \cr
 0.125\times S^{-0.9}         &  $S < 0.015~Jy$ \cr
 }
\label{eq.L1}
\end{equation}

Once the LogN-LogS of a population of sources is known in a given energy band it is
possible to estimate their emission in other parts of the electromagnetic spectrum,
provided that the flux ratio between the two bands is known. In this section we deal with flux ratios
and Spectral Energy Distributions (SED) of blazars that will be used to estimate the contribution to
cosmic backgrounds at microwave frequencies and in the X-ray and $\gamma$-ray bands.

\begin{figure}[ht]
\vbox{ \centerline{
\includegraphics[width=5.cm, angle=0] {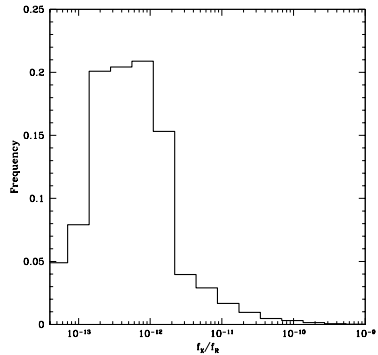}
}}
\caption{The \fxfr distribution of blazars estimated by \cite{Giommi06}}
 \label{fxfr}
\end{figure}

The distribution of \fxfr flux ratios shown in Fig.\ref{fxfr}  (see \cite{Giommi06} for details) spans about four orders of magnitudes implying that the X-ray flux of blazars with the same radio flux can be different by up to a factor 10,000!
\begin{figure}[ht]
\vbox{ \centerline{
\includegraphics[width=5.truecm, angle=-90] {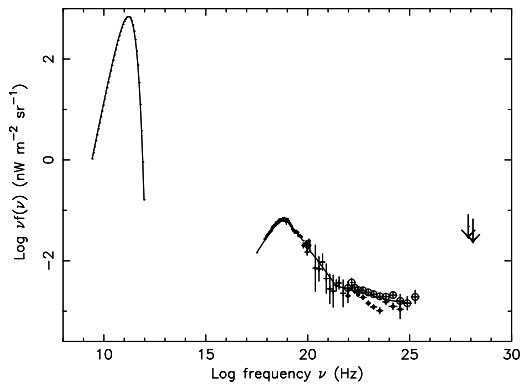}
}}
 \caption{The extragalactic Cosmic Background Energy distribution at microwave, X-ray
and $\gamma$-ray energies.
The CMB is represented by a black-body spectrum with temperature of 2.725 $^o$K \cite{Mather99}; the CXB is taken from HEAO-1 measurements \cite{Marshall80,Gruber99} and has been scaled to match
the more recent \sax, ASCA, and XMM-Newton results in the 2-10 keV band
\cite{Vecchi99,Lumb02,Kushino02}. The gamma ray background is derived from the COMPTEL
data in the range $0.8 - 30$  MeV \cite{Kappadath98} and from EGRET data in the range
$30$ MeV - $50$ GeV.   We report the results of two different
analyses of the EGRET data: open circles from \cite{Sreekumar98} and filled circles from
\cite{Strong04}, which uses an improved model of the Galactic diffuse continuum
gamma-rays. As for the TeV diffuse background, we report the upper limits in the $20$ -
$100$ TeV region derived from the HEGRA air shower data analysis \cite{hegra01}.
In the 1 TeV -- 1 PeV energy range, other experiments give only upper limits and there is no clear observation of a diffuse photon signal yet (see \cite{Giommi06} for more detail).
}
 \label{CBGs}
\end{figure}

In the following we estimate the blazar contribution to the cosmic backgrounds  basing our calculation on the radio LogN-LogS of Fig. \ref{logns} and on flux ratios in different bands (Fig \ref{fxfr}) or on  observed blazar SED.

\subsection{The Cosmic Microwave Background}

The contribution of blazars to the CMB has been estimated in the past from different
viewpoints (e.g. \cite{toff98,giocol04}).  Here  we calculate the integrated microwave background intensity as follows
\begin{equation}
 \displaystyle {I_{blazars}=\int_{0.1mJy}^{1Jy} S~{dN \over dS}~dS } ~,
 \label{eq.IBLazars}
\end{equation}
where dN/dS is the differential of eq. (\ref{eq.L1}). The minimum integration flux of 0.1
mJy for $S_{min}$ is likely to be conservative, since blazars with radio flux near or
below 1 mJy are already included in the {\it Einstein} Medium Sensitivity Survey BL Lac
sample \cite{Rec00}. The integrated intensity $I_{blazars}$ is  converted from 5GHz to
microwave frequencies by convolving the flux value with the observed distribution of
spectral slopes between 5GHz to microwave frequencies, as estimated from the 1Jy-ARN
sample (see  \cite {giocol04} for details).

\begin{figure}[ht]
\hbox{ \centerline{
\includegraphics[width=5.cm, angle=0] {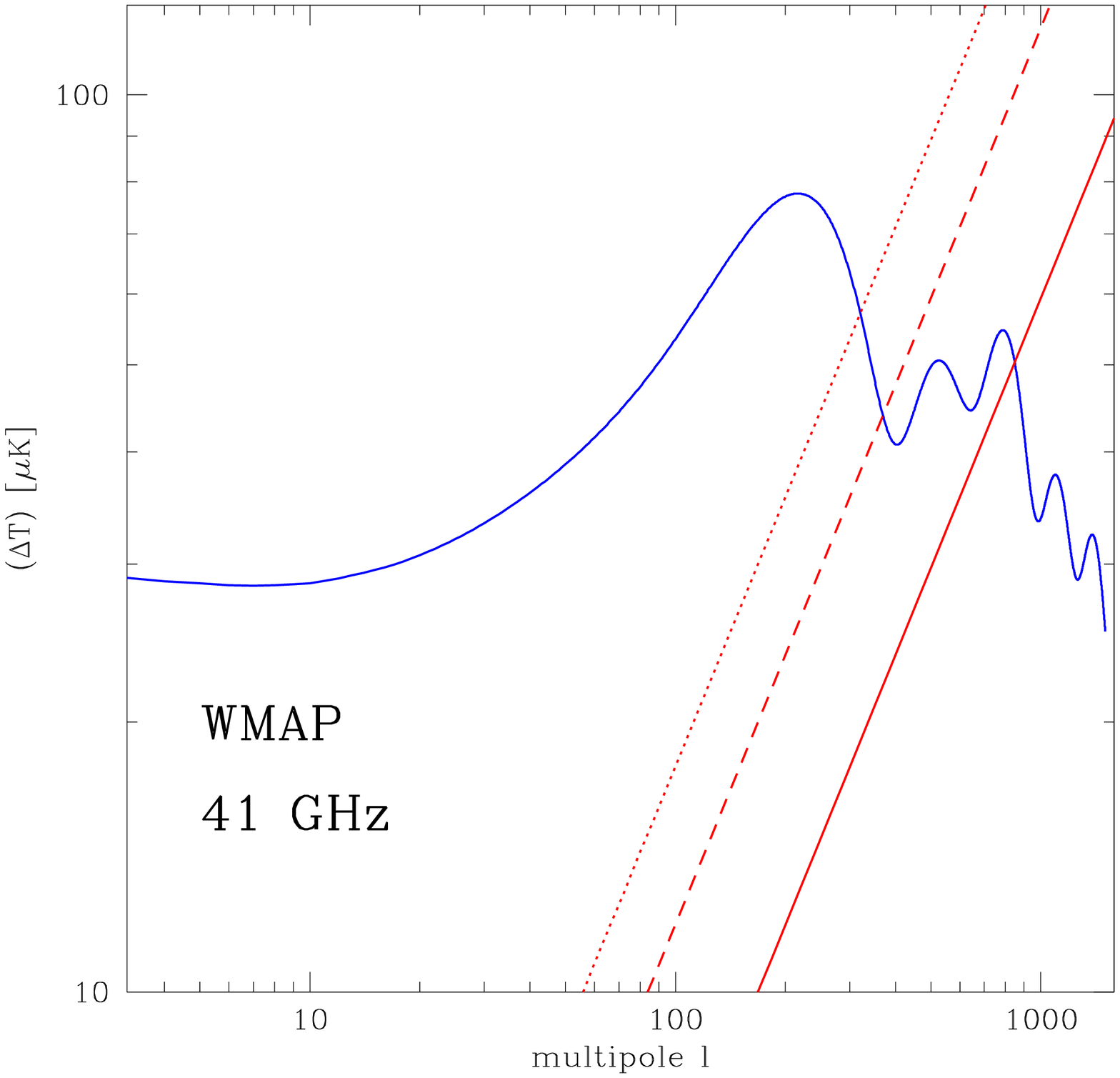}
\includegraphics[width=5.cm, angle=0] {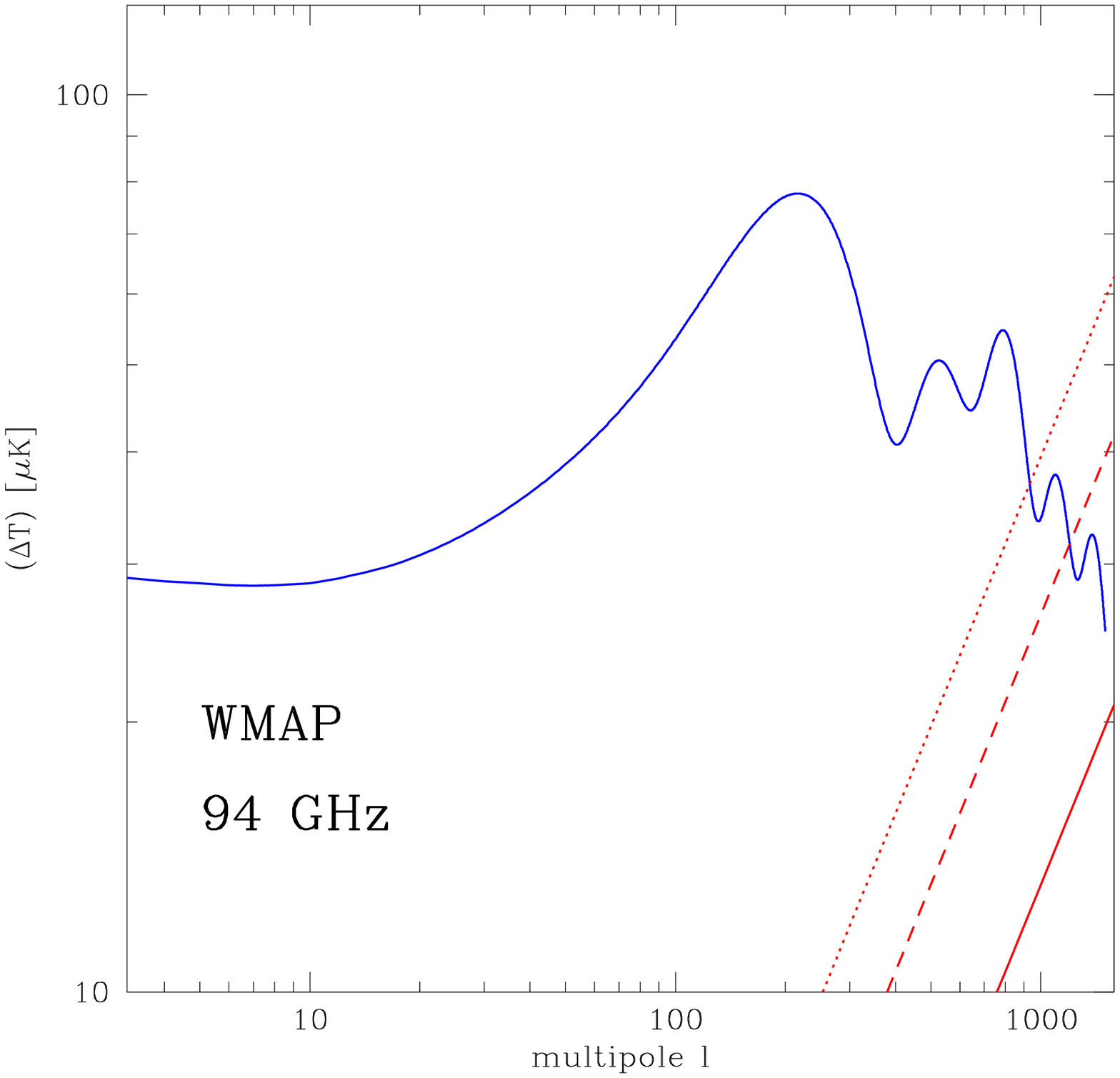}
}}
 \caption{The contribution of blazars to the CMB fluctuation spectrum in the WMAP
41GHz (left panel) and 94 GHz (right panel) channels, as evaluated from the
LogN-LogS given in Fig. \ref{logns} (solid line).
We also show the angular power spectrum for the blazar population by adding an estimate
of the possible contribution of radio sources with steep-spectrum at low radio
frequencies which flatten at higher frequencies (dashed line). The dotted line  also
includes the effect of spectral and flux variability. Although
this additional contamination may be substantial a precise estimation can only be done
through simultaneous high resolution observations at the same frequency. A typical CMB
power spectrum evaluated in a $\Lambda$CDM cosmology with $\Omega_m=0.3,
\Omega_{\Lambda}=0.65, \Omega_b=0.05$ which best fits the available data is shown for
comparison. }
 \label{cmb41_94}
\end{figure}
The fractional contamination, $I_{blazars}/I_{CMB}$ of blazars to the CMB intensity ($I_{CMB}$) and the corresponding apparent temperature increase at frequency of 90-300 GHz has been derived by
\cite{Giommi06}.  Blazars also contribute to the temperature anisotropy spectrum of the CMB with a spectrum
\begin{center}
 $(\Delta T)_{blazar}= [(2 \pi)^{-1} C_{\ell} \ell (\ell +1)]^{1/2}$, where
 \end{center}
 \be
 C_{\ell, {\rm blazar}} = \int_{S_{\rm min}}^{S_{\rm max}} dS~{dN \over dS}~S^2 ~,
 \label{Eq.cl}
 \ee
 The quantity on the right hand side in eq. (\ref{Eq.cl}) is the usual Poisson shot-noise term, \cite{Peebles80,Tegmark96} and we neglect here the clustering term, $\omega(I)^2$, which adds to the Poissonian term since there is not a clear estimate of the blazar clustering  yet. Note, however, that the
inclusion of clustering significantly increases the amount of contamination,
especially at large angular scales \cite{Gonzales05}.

For the blazar population described by the LogN-LogS given in Fig. \ref{logns}, we found
$C_{\ell, {\rm blazar}} \approx 54.4 ~ {\rm Jy}^2 {\rm sr}^{-1}$ at 41 GHz and $C_{\ell,
{\rm blazar}} \approx 51.6~{\rm Jy}^2 {\rm sr}^{-1}$ at 94 GHz. When translated into temperature
units (see \cite{giocol04}) we found a value $C_{\ell, {\rm blazar}} \approx 2.22\cdot 10^{-2} ~\mu {\rm K}^2 sr$ at
41 GHz and $C_{\ell, {\rm blazar}} \approx 1.09 \cdot 10^{-3} ~\mu {\rm K}^2 sr$ at 94
GHz. We show in Fig. \ref{cmb41_94}  the quantity ($\Delta T$)$_{\rm blazar}$
and compare it to the CMB fluctuation power spectrum which best fits the available data.
These values (shown as solid lines) have to be considered as a definite lower
limit for $C_{\ell,{\rm blazar}}$, since they neglect the contribution of steep-spectrum
sources at low frequencies which flatten at these frequencies (41 and 94 GHz, dashed lines)
and the effect of flux variability (dotted lines).

The blazar flux variability at millimeter wavelengths may be substantial (higher than factors 3-10
on time scales of weeks to years seen at cm wavelengths) and could increase the amount of
contamination of CMB maps when these are built over long integration periods.

The contamination level shown in  Figs. \ref{cmb41_94} is the one expected
in the case where no blazars are removed from the CMB data. We expect that a correct procedure
to derive the CMB power spectrum that takes into account the full point-like source contribution implied by our LogN-LogS, would both influence the shape of the expected power spectrum
and increase the statistical uncertainties of the WMAP data points, especially at high
multipoles, where the blazar contribution is larger.

The previous calculations we performed neglecting the blazar clustering and thus they
must  again be considered as a lower limit to the realistic angular power spectrum contributed by these sources. The effects of clustering on the CMB fluctuation spectrum has been partially estimated by some authors: (e.g. \cite{Gonzales05,Scott99}).
Expectations for the clustering effect strongly depend both on the adopted model for the
source counts and on their clustering model. Based on the correlation function of
\cite{Loan97} on that for the SCUBA sources \cite{Scott99}, and on our blazar LogN-LogS, we expect that the contamination of the
first peak of the fluctuation spectrum (at the WMAP 41 GHz channel) is at a level in the
range 20-25 \%. This estimate does not include possible variability effects and
additional core-dominated radio sources

\subsection{The Soft X-ray background}

We estimate  the contribution of blazars to the CXB at 1 keV using two methods: i)
converting the integrated radio flux calculated with Eq. (\ref{eq.IBLazars}) into X-ray
flux with the observed distribution of \fxfr flux ratios in Fig. \ref{fxfr}; and ii)
converting the integrated blazar contribution to the CMB (at 94 GHz) to X-ray flux using
the distribution of \amx, the microwave (94 GHz) to X-ray (1keV) spectral slope,
estimated from the subsample of blazars detected by WMAP in the 94 GHz channel, for
which an X-ray measurement is available (see Fig. 9 in \cite{Giommi06}).
The first method gives a total blazar contribution to the X-ray background of $2.7\times
10^{-12}$ \ergssd (about 70\% of which is due to HBL sources with \fxfr  $> 5 \times
10^{-12}$ \ergj) in the ROSAT 0.1-2.4 keV energy band. Assuming \cite{Giommi06} an
average blazar X-ray energy spectral index of \ax = 0.7, this flux converts to $2.6\times
10^{-12}$ \ergssd in the 2-10 keV band or 11\% of the CXB, which is estimated to be
$2.3\times 10^{-11}$ \ergssd \cite{perri00}.

The distribution of ${\alpha_{\mu x}}$ has an average $\langle$~\amx~$\rangle$ =  1.07
and a standard deviation of 0.08 corresponding only to about a factor 3 in flux. This
distribution is much narrower than the one between the radio and X-ray band (Fig.
\ref{fxfr}), while the dispersion is comparable to that expected from blazar variability
at radio and especially at X-ray frequencies. The X-ray flux can therefore be estimated
simply as $f_{1keV}$=1.4$\times 10^{-7}~f_{94GHz} $ (see, e.g., \cite{Giommi06}).

Since the blazar integrated emission at 94 GHz is $7.2\times10^{-6}$CMB$_{94GHz}$ or 0.64
Jy/deg$^2$ and the cosmic X-ray background is $2.3\times 10^{-11}$ \ergs \cite{perri00}
in the 2-10 keV band (equivalent to 2.31$\mu$Jy/deg$^2$ at 1 keV), we have that f$_{1keV}$ = 0.09 $\mu$Jy/deg$^2$ or 3.9\% of the CXB for $f_{94GHz}$=0.64Jy/deg$^2$ .
Considering that the ${\alpha_{\mu x}}$ distribution of \cite{Giommi06} only includes LBL objects and that
HBL sources make up two thirds the total contribution to the CXB, the percentage scales to
about 12\% which is very close to the 11\% obtained with the previous method.
Both results are in good agreement with the independent estimate of \cite{Galbiati04},  who find  that the radio loud AGN content of the CXB is 13\% in the XMM-Newton bright serendipitous survey.

\subsection{Hard X-ray -- soft $\gamma$-ray Background}

The number of sources detected at energies larger than soft X-rays is still rather low,
so building reliable distributions of flux ratios between radio or microwaves and the
Hard X-ray/$\gamma$-ray fluxes is not currently possible. Thus, in order to estimate the
blazar contribution to high energy Cosmic Backgrounds (E$~>~$100 keV), we therefore
followed a different approach: we extrapolated the predicted blazar integrated intensity
at microwave frequencies (eq. \ref{eq.IBLazars}) to the hard X-ray and soft $\gamma$-ray
band using a set of hypothetical SSC spectral energy distributions.

Figure \ref{hard_backs} (left panel) shows the CMB, CXB, and CGB together with three predicted SEDs
from a simple homogeneous SSC models whose parameters are constrained to 1) be consistent
with the expected integrated flux at 94 GHz, 2) have the \amx slope equal to the mean
value of the WMAP blazars (\amx = 1.07), and 3) possess a radio spectral slope equal to
the average value in the WMAP sample. The three curves, so constrained, are characterized
by synchrotron peak frequencies of \nupeak = $10^{12.8}$, $10^{13.5}$, and $10^{13.8}$
Hz. A high value of \nupeak largely overestimates the observed hard-X-ray to soft
$\gamma$-ray ($\approx 10^{20}-2\times10^{22}$ Hz or $\approx$ 500 keV-10 MeV) cosmic
background, whereas a too low value of \nupeak predicts a negligible contribution (see
Fig. \ref{hard_backs}, left panel). The case \nupeak = $10^{13.5}$ Hz predicts 100\% of the cosmic
background. Since the Log(\nupeak) values of blazars in the WMAP and other catalogs peak
near 13.5 and range from 12.8-13.7 within one sigma from the mean value, we
conclude that blazars may be responsible for a large
fraction (possibly 100\%) of the hard-Xray/soft $\gamma$-ray cosmic background.

\subsection{$\gamma$-ray Background}

\begin{figure}[h]
\begin{center}
\includegraphics[height=11.cm, width=4.5cm, angle=-90] {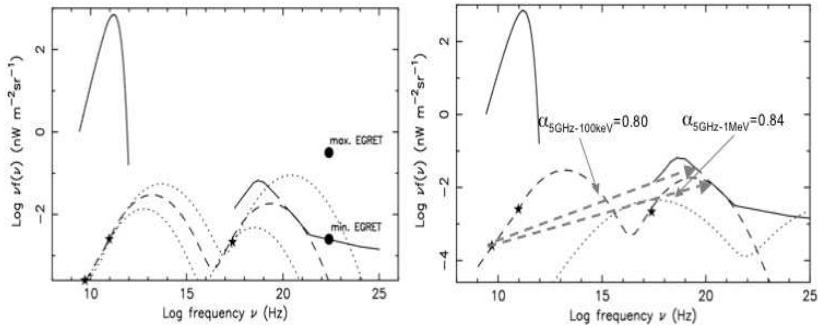}
\end{center}
 \caption{Left: The possible contribution of LBL blazars to the  cosmic
background. Simple SSC energy distributions fail to reproduce the observed
intensity and slope of the $\gamma$-ray background. The observed range of  $\gamma$-ray
emission with EGRET, normalized to radio flux,  indicates that the duty cycle at these
energies must be low (see text for details). Right: The spectral slopes between the 5GHz emission and the 100 keV and 1 MeV high energy fluxes in the SSC distribution of LBL blazars that reproduces
the observed X and $\gamma$-ray cosmic background.
 }
\label{hard_backs}
\end{figure}

The SSC distributions of Fig. \ref{hard_backs} predict a negligible blazar contribution
to the extragalactic $\gamma$-ray Background above $100~$MeV. Nevertheless, it is well
known that blazars are the large majority of the extragalactic $\gamma$-ray (E $>$ 100
MeV) identified sources detected by the EGRET experiment \cite{Hartman99}; therefore they
are likely to contribute to the $\gamma$-ray background in a significant way. Indeed,
\cite{P93} concluded that blazars should make a large fraction, if not the totality, of
the extragalactic $\gamma$-ray background. However, these early calculations relied
upon a very small sample and had to assume no strong variability, a characteristic that was later demonstrated to be extremely common in $\gamma$-ray detected blazars.

Figure 11  in \cite{Giommi06}  compares the energy distribution of the CMB, CXB, and CGB
to the SED of 3C 279, a well-known bright blazar  detected by EGRET, scaled to the
integrated blazar flux intensity as calculated with eq. (\ref{eq.IBLazars}). Taking into
account the strong variability of blazars seen across the whole e.m. spectrum
\cite{Giommi06} show that while the contribution to the CXB can range from a few percent
to over 10\% in the higher states, the predicted flux at $\gamma$-ray frequencies ranges
from about 100\% to several times the observed cosmic background intensity. Such large
excess implies that either sources like 3C279 are not representative of the class of
blazars or their duty cycles at $\gamma$-ray frequencies are very low. A way of
quantifying the ratio between the  $\gamma$-ray intensity predicted by assuming that the
source is  representative of the entire population and the actual background intensity is
to use the microwave (94 GHz) to $\gamma$-ray (100 MeV) slope \amg =
-$Log(f_{94GHz}/f_{100MeV})\over{Log(\nu_{94GHz}/\nu_{100MeV}})$.

For comparison, we defined \amgB as the \amg of a hypothetical source that would produce
100\% of the CGB if representative of the entire class. Any real source with \amg flatter
than \amgB would then predict an integrated flux in excess of the observed $\gamma$-ray
background if representative of the entire population; alternatively, its duty cycle must
be lower than 100\%.  Fig. \ref{hard_backs} (left) shows the minimum and maximum values of \amg among the sources detected by the EGRET experiment.
\subsection{TeV background}

All blazars detected  so far at TeV energies are of the HBL type, therefore we estimate the blazar contribution to the TeV background as in the case of the soft $\gamma$-ray Background but only considering the HBL component and not the entire blazar population. These extreme objects have been estimated to be  about 0.1 \% of the blazar population \cite{Giommi06}.
Using the SED of the well known TeV blazar MKN421 normalized at 94 GHz (see Fig.13 in
\cite{Giommi06}), so that the flux is scaled to 0.1\% of the intensity produced by the
entire population of blazars, we see that, despite HBLs are a tiny minority, their
integrated X-ray flux makes up a fairly large fraction of the CXB and that their TeV emission may
produce a significant amount of extragalactic light.
We note, however, that since extreme HBLs, such as those of the Sedentary survey, are
very rare (one object in several thousand degrees with flux above a few mJy), the
extragalactic light at TeV energies should be very patchy, associated to single sources,
rather than a diffuse light resulting from the superposition of many unresolved discrete
sources.

\section{Prospects for future high-energy observations of blazars}

We have seen that blazars, and in general NTED-AGN, emit non-thermal radiation across the entire
electromagnetic spectrum. A clear understanding of the blazar content of the extragalactic high-energy sky is necessary as new missions are planned or are studied by European and other space agencies to explore in more detail the hard X-ray and  $\gamma$-ray bands and to open the last energy windows (e.g. the few hundred keV-few MeV region) that for technical difficulties have so far remained almost unexplored but that are extremely important for the understanding of the physical processes powering NTED-AGN.

In the following we apply the same method used in the previous sections to estimate the
number of blazars that could be detected by future microwave, X-ray and $\gamma$-ray
observatories.
We report the results in Table \ref{tab.predictions} where column 1 gives the
energy band considered, column 2 gives the limiting sensitivity reached in that energy band,
column 3 gives the corresponding 5GHz flux, and column 4 gives
the number of blazars expected in the high Galactic latitude sky ($\approx 25,000
deg^2$).

Figure \ref{fxfr} shows that the distribution of \fxfr ratios is extremely broad
reflecting the fact that the flux in the X-ray region is deeply affected by the position
of the synchrotron peak in  $Log(\nu)-Log(\nu f(\nu))$ plots.  Blazars where the peak is located in
the far infrared (LBL) have very little X-ray flux whereas extreme HBL objects can be
thousands of times brighter at X-ray frequencies than LBLs of the same radio flux. Figure
\ref{fxfr} implies that a survey as deep as  $10^{-15}$ \ergs would detect the whole
\fxfr distribution of blazars with 5GHz flux larger than 10 mJy, corresponding to well
over 100,000 sources in the high Galactic latitude sky.
From Fig. \ref{hard_backs} (right panel) we can see that the average slope between the
5GHz radio flux and the hard-X-ray band and the soft $\gamma$-ray band is  0.85 and 0.84
respectively. Since the ratio between the radio flux and these energy bands is not
expected to have a large spread around the mean value like \fxfr, the average slopes can be directly
converted to predict the hard-X-ray and the soft $\gamma$-ray fluxes.   We have reported
these predictions in column 2 of Table \ref{tab.predictions}
\begin{table}
\begin{center}
\caption{The number of blazars in the high Galactic latitude sky at different
sensitivities and energy bands}
\label{tab:1}       
\begin{tabular}{llll}
\hline\noalign{\smallskip}
 Energy  band  & Flux limit &Flux limit & No. of sources   \\
    & (energy band) & (5GHz)& at high b$\|$    \\
        &  && (25,000 $deg^2$)   \\
 \noalign{\smallskip}\hline
 \noalign{\smallskip}
 Micro-wave  &70 mJy&100 mJy&  ~~~~~~5000\\
  (100~GHz)    &~7 mJy&~10 mJy&  $ >$100,000\\
 \noalign{\smallskip}\hline
 Soft X-Ray  &$10^{-15}$ \ergs &$\sim$0.1-10 mJy& $ >$100,000  \\
   (0.1$-$2.4~keV)  &  &  \\
 \noalign{\smallskip}\hline
 Hard X-Ray  &$5~10^{-12}$ \ergs &1000 mJy& ~~~~~~~~130  \\
   $50-150$~keV &$1~10^{-13}$ \ergs &~~~20 mJy & ~~$ \approx$70,000 \\
 \noalign{\smallskip}\hline
 Soft $\gamma$-ray  &$5~10^{-7}$ \phs &1000 mJy&  ~~~~~~~~130 \\
  $3-10$MeV & $1~10^{-7}$ \phs & ~~200 mJy&~~~~~~~2000\\
 \noalign{\smallskip}\hline
 $\gamma$-ray   &$1~10^{-7}$ \phs &Monte Carlo& $~~~~\approx 100$ \\
  $100~$MeV-$100~$GeV  &$3~10^{-9}$ \phs &simulation& $~~~\approx 3000$\\
 \noalign{\smallskip}\hline
\end{tabular}
\end{center}
\label{tab.predictions}
\end{table}

\noindent {\bf A deep all sky survey in the soft X-ray band: 0.1-10 keV.}  From Table  \ref{tab.predictions}
we see that  a deep all-sky X-ray survey in the soft X-ray
band would allow us to build a sample of $>$ 100,000 blazars that are expected to be above the limiting sensitivity of PLANCK.
Given the significant impact of the blazar foreground emission on the CMB power spectrum it is
 important to remove this contaminating component from the CMB as much as possible.
An efficient way to achieve this is to exploit the fact that the spectral slope distribution between microwave and soft X-ray flux of LBLs is very narrow (see Fig. 9 of \cite{Giommi06}) with a dispersion that is probably mostly due to intrinsic variability. The soft X-ray flux of LBLs (that is
$>$90\% of the blazar population) is therefore a very good estimator of the flux at
microwave frequencies and could be used to locate and remove foreground blazars from the
PLANCK and other CMB missions.

The extremely large sample of blazars
produced by such a survey could also be used to study the statistical properties of
blazars in great detail, including the spatial correlation function, and would identify a
very large number of targets for the next generation of $\gamma$-ray observatories such
as AGILE, GLAST, and future instruments operating in the still poorly explored MeV
spectral region.

The expected number of blazars at $\gamma$-ray energies above 100 MeV cannot be estimated
with a simple extrapolation of the 5GHz flux (see Sect. 2.4) and therefore we have
calculated the preliminary expectations listed in the last row of Table
\ref{tab.predictions} through a Monte Carlo method that makes use of the radio luminosity
distribution of blazars and predicts $\gamma$-ray fluxes taking into account the \amg
distribution of EGRET detected blazars. This method is currently under development as part of
our contribution to the GLAST project.

\begin{acknowledgements}
It is a pleasure to thank the Organizing and Advisory Committee of this meeting for the
invitation to present our results in the context of the planned and future gamma-ray
experiments.
\end{acknowledgements}



\end{document}